\documentclass[num-refs]{wiley-article}
\pdfoutput=1
\usepackage{amsmath}
\usepackage{amssymb}
\usepackage{amsfonts}
\usepackage{subcaption}
\usepackage{algorithm}
\usepackage{algpseudocode}
\usepackage[utf8]{inputenc}
\usepackage[table]{xcolor}
\algnewcommand{\Initialize}[1]{%
  \State \textbf{Initialize:}
  \Statex \hspace*{\algorithmicindent}\parbox[t]{.8\linewidth}{\raggedright #1}
}
\papertype{REVIEW PAPER}
\paperfield{Magnetic Resonance in Medicine}
\title{CG-SENSE revisited: Results from the first ISMRM reproducibility challenge}
\author[1]{Oliver~Maier}
\author[2]{Steven~H.~Baete}
\author[3]{Alexander~Fyrdahl}
\author[4,5]{Kerstin~Hammernik}
\author[6]{Seb~Harrevelt}
\author[7,8\authfn{2}]{Lars~Kasper}
\author[9]{Agah~Karakuzu}
\author[10]{Michael~Loecher}
\author[7]{Franz~Patzig}
\author[11,12]{Ye~Tian}
\author[13]{Ke~Wang}
\author[14]{Daniel~Gallichan}
\author[15,16,17,18]{Martin~Uecker}
\author[2]{Florian~Knoll}

\affil[1]{Institute of Medical Engineering, Graz University of Technology, Graz, Austria}
\affil[2]{Center for Biomedical Imaging, New York University School of Medicine, New York, New York, USA}
\affil[3]{Department of Clinical Physiology, Karolinska University Hospital, and Karolinska Institutet, Stockholm, Sweden}
\affil[4]{Department of Computing, Imperial College London, London, United Kingdom}
\affil[5]{Institute of Computer Graphics and Vision, Graz University of Technology, Graz, Austria}
\affil[6]{Department of Biomedical Engineering, Eindhoven University of Technology, Eindhoven, The Netherlands}
\affil[7]{Institute for Biomedical Engineering, ETH Zurich and University of Zurich, Zurich, Switzerland}
\presentadd[\authfn{2}]{Techna Institute, University Health Network, Toronto, Canada}
\affil[8]{Translational Neuromodeling Unit, Institute for Biomedical Engineering, University of Zurich and ETH Zurich, Zurich, Switzerland}
\affil[9]{NeuroPoly Lab, Institute of Biomedical Engineering, Polytechnique Montr\'eal, Montr\'eal, Canada}
\affil[10]{Department of Radiology, Stanford University, Stanford, California, USA}
\affil[11]{Utah Center for Advanced Imaging Research (UCAIR), Department of Radiology and Imaging Sciences, University of Utah Salt Lake City, UT, USA}
\affil[12]{Ming Hsieh Department of Electrical and Computer Engineering, Viterbi School of Engineering, University of Southern California, Los Angeles, California, USA}
\affil[13]{Department of Electrical Engineering and Computer Sciences, University of California, Berkeley, Berkeley, California, USA}
\affil[14]{Cardiff University Brain Research Imaging Centre, Cardiff, UK}
\affil[15]{Institute for Diagnostic and Interventional Radiology, University Medical Center G\"ottingen, G\"ottingen, Germany}
\affil[16]{German Centre for Cardiovascular Research (DZHK)}
\affil[17]{Cluster of Excellence "Multiscale Bioimaging: from Molecular Machines to Networks of Excitable Cells" (MBExC), University of G\"ottingen, Germany}
\affil[18]{Campus Institute Data Science (CIDAS), University of G\"ottingen, Germany}
\corraddress{Florian Knoll, Center for Biomedical Imaging, New York University School of Medicine, New York, New York, USA}
\corremail{florian.knoll@nyumc.org}
\fundinginfo{Austrian Academy of Sciences, DOC-Fellowship: 24966; NIH (NIBIB) awards R01EB024532, P41EB017183, R21EB027241, U24EB029240.}
\runningauthor{Oliver Maier et al.}
\begin{document}
\maketitle
\vspace{5pt}
{\bf WORD COUNT:} Approximately 7400
\newpage
\begin{abstract}
\textbf{Purpose} 

The aim of this work is to shed light on the issue of reproducibility in MR image reconstruction in the context of a challenge. Participants had to recreate the results of "Advances in sensitivity encoding with arbitrary k-space trajectories" by Pruessmann et al.

\textbf{Methods} 

The task of the challenge was to reconstruct radially acquired multi-coil k-space data (brain/heart) following the method in the original paper, reproducing its key figures. Results were compared to consolidated reference implementations created after the challenge, accounting for the two most common programming languages used in the submissions (Matlab/Python).

\textbf{Results} 

Visually, differences between submissions were small. Pixel-wise differences originated from image orientation, assumed field-of-view or resolution. The reference implementations were in good agreement, both visually and in terms of image similarity metrics.

\textbf{Discussion and Conclusion}

While the description level of the published algorithm enabled participants to reproduce CG-SENSE in general, details of the implementation varied, e.g., density compensation or Tikhonov regularization. Implicit assumptions about the data lead to further differences, emphasizing the importance of sufficient meta-data accompanying open data sets.

Defining reproducibility quantitatively turned out to be non-trivial for this image reconstruction challenge, in the absence of ground-truth results. Typical similarity measures like NMSE of SSIM were misled by image intensity scaling and outlier pixels. 

Thus, to facilitate reproducibility, researchers are encouraged to publish code and data alongside the original paper.
Future methodological papers on MR image reconstruction might benefit from the consolidated reference implementations of CG-SENSE presented here, as a benchmark for methods comparison.
%
\end{abstract}
\newpage
\normalsize
\section{Introduction}

\begin{figure}
    \centering
    \includegraphics[width=\textwidth]{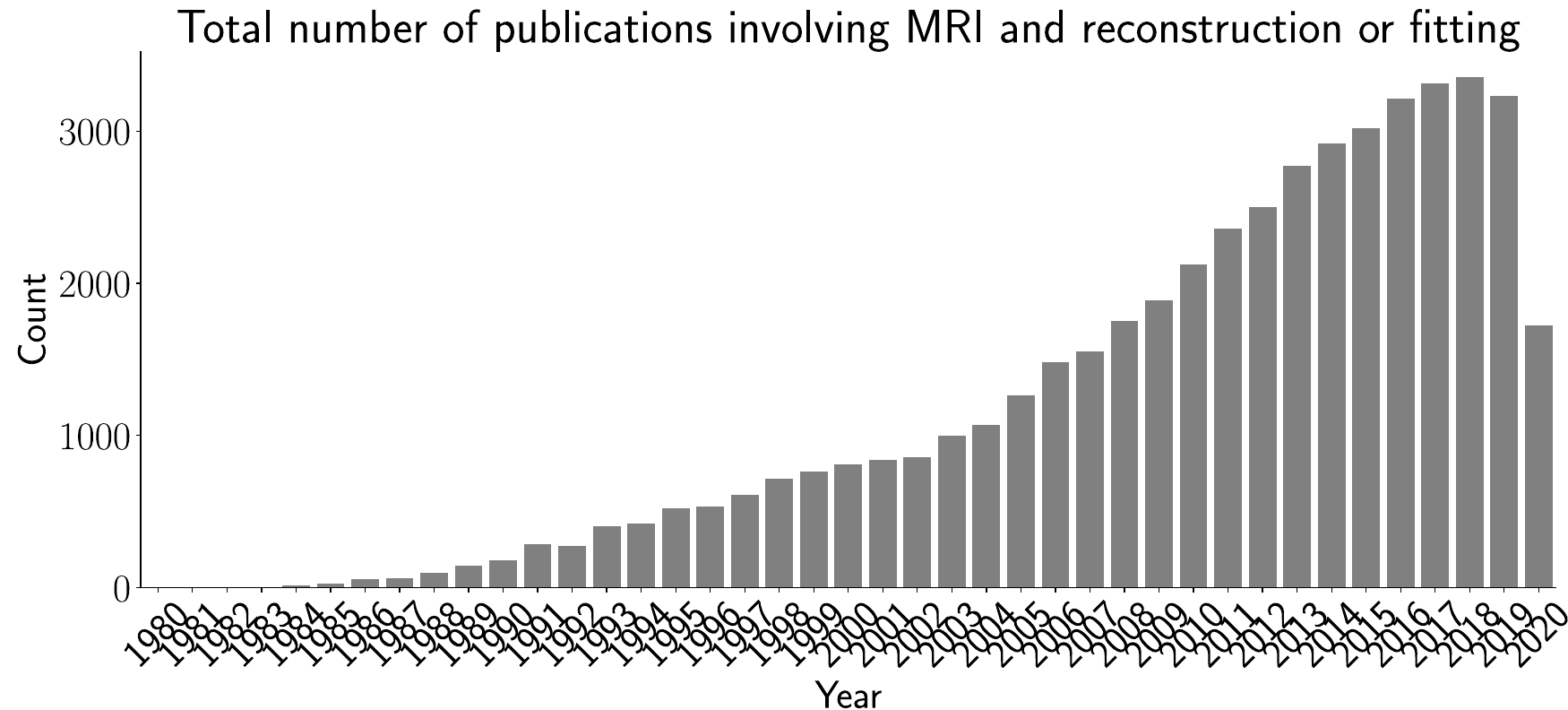}
    \caption{Number of publications on PubMed including "MRI" and either 
"reconstruction" or "fitting". Data search done on the 4$^{\text{th}}$ of 
August, 2020.}
    \label{fig:papers}
\end{figure}

Over the past decades MRI experienced a vast thrust towards an algorithmic perspective owing to the increased computational power of standard computers leading to the invention and development of numerous reconstruction methods. This is reflected in the tremendous increase of publications registered on Pubmed that involve ‘MRI’ and either ‘reconstruction’ or ‘fitting’ over the last 2 decades (see Figure~\ref{fig:papers}). The peak of 3354 publications in 2018 amounts to an average of 9 papers per day. Typically, computational innovation in these papers is shown by comparing novel methods to established algorithms in the field via suitable quality metrics.

One of the fundamental computational approaches to image reconstruction is parallel imaging, i.e. the idea to use a-priori knowledge about multiple receiver coil sensitivities to accelerate scans. Image reconstruction then shifts from a simple Fourier transform – which may optionally include a gridding step for non-Cartesian data – to solving a more complex inverse problem, based on a matrix equation of image encoding, as proposed in a general form in Pruessmann et al. ~\citet{pruessmann2001} commonly referred to as "conjugate gradient CG-SENSE". A lot of image reconstruction papers published thereafter refer to this standard algorithm, often performing direct comparisons to prove the efficacy of their method. However, no commonly agreed-on reference implementation of the CG-SENSE algorithm is readily available. Therefore, these comparisons to the ‘SENSE’ method are mere comparisons to one version of it, be it custom implementations, those based on openly available image reconstruction toolboxes, or even obtained from a black-box implementation provided by the scanner vendors. This lack of a refererence implementation reflects a fundamental problem of research reproducibility in the MR image reconstruction domain.

The reproducible research study group (RRSG) of the ISMRM aims to enhance reproducibility by facilitating fair and simple comparisons to existing algorithms. However, comparing novel algorithms to re-implementations of published work without having access to the original code can lead to wrong conclusions. Often, algorithmic details are not reported in detail in publications and small deviations of input parameters can lead to strong differences in the output, regularly degrading the performance of the existing method, which is a general problem faced in the scientific community~\citep{claerbout1992, represearch2010, peng2011, baker2016, topalidou2016, chen2018, cacho2020}. To that end, the RRSG announced a reproducibility challenge in April 2019 as part of the Annual Meeting of the ISMRM in Montreal. The goal was to reproduce the core findings of the paper "Advances in sensitivity encoding with arbitrary k-space trajectories" from~\citet{pruessmann2001}], solely based on the description available in the paper, and to converge towards a reference implementation being accessible to the community. Participants were required to reproduce the main figures of the original paper given two fully-sampled radial brain and heart datasets. Signal and trajectory data were supplied but neither sensitivity maps nor noise covariance scans. No programming language restrictions were given, as long as the source code was shared and the computational results could be reproduced. The detailed instructions can be found in the corresponding ISMRM blog post~\footnote{https://blog.ismrm.org/2019/04/02/ismrm-reproducible-research-study-group-2019-reproduce-a-seminal-paper-initiative/}.


In this work, we present the outcome of this initiative, compare the different submissions and discuss potential problems in reproducing the findings of a scientific paper solely from the manuscript. Furthermore, we consolidated the submissions from the participating groups into two reference implementations (Python and Matlab), which are available online in the ISMRM git repository and could serve as a benchmark for future publications seeking comparison to CG-SENSE. The reference implementations will be discussed in more detail in section~\ref{sec:refimplement}, specifically focusing on critical points in the implementation. The main features of each submission will be shown in section~\ref{sec:submissions} and differences regarding implementation details and possible sources of deviations to the reference implementations will be discussed. Finally, recommendations for conducting reproducible research and future challenges are given.


\section{Methods}
\subsection{Design of the first RRSG challenge}
Since this was the first ever reproducibility challenge by the study group, we designed it around a rather simple premise to encourage submissions from the community. This started with the choice of the paper. We wanted a paper that is seminal in our field, where the authors did not already provide a reference implementation themselves. We wanted to be able to provide all the data ourselves and not rely on any closed source or proprietary software for any step of the data processing. We also wanted a paper where we expected the results of the challenge to be uncontroversial. In fact, we expected that the submissions of the participants would successfully reproduce the main results of the paper without showing fundamental differences, but still would reveal some interesting differences that we could learn from about reproducibility issues. Finally, since one of the goals of this initiative is to build up a library of standard reference implementations that can be used for comparison when publishing new methods, we wanted to cover a method that is commonly used as a reference by MRM authors. A second design choice was the timeline. We wanted the turnaround of the participants to be relatively quick, because we wanted to see how well the paper can be reproduced within a time-frame of a couple of weeks. In particular, we announced the challenge and provided the material on March 28th 2019, and set the deadline for submissions for May 1st 2019.

In the rest of this section, we are providing a brief review of the CG-SENSE method that was introduced in~\citet{pruessmann2001}, a detailed description of the data that was used for the challenge and an overview of the individual submissions and finally a description of the consolidated reference implementations that were developed after the conclusion of the challenge. 

\subsection{CG-SENSE}
Throughout this work let $n_x\times n_y$ denote the image dimension in pixels and $n_c$ the number of receive coils. For simplicity, assume that $n_x = n_y = n$. The total number of k-space samples, i.e. number of read-outs times number of read-out points, is denoted as $n_k$. Reconstructing an image $\mathbf{v}\in\mathbb{C}^{n^2}$ from undersampled data $\mathbf{m}\in\mathbb{C}^{n_k\,n_c}$ acquired with multiple receive coils $n_c$ is an inverse problem following a linear encoding process
\begin{equation}
    \mathbf{Ev} = \mathbf{m}
    \label{eq:lin}
\end{equation}
with $\mathbf{E}:\mathbb{C}^{n^2}\to\mathbb{C}^{n_k\,n_C}$ being the linear encoding matrix, mapping from image space to k-space~\citep{pruessmann1999, pruessmann2001}. 
The encoding matrix $\mathbf{E}$ describes the whole MRI acquisition pipeline, consisting of coil sensitivity profiles $S_\gamma\in\mathbb{C}^{n_k\,n_C}$ and Fourier transformation combined with the sampling operator, i.e. the non-uniform FFT ($\text{NUFFT}: \mathbb{C}^{n_k\,n_C}\to\in\mathbb{C}^{n^2}$). Assuming Gaussian noise in the acquired k-space data, the optimal solution regarding the signal-to-noise ratio (SNR) is given by the minimum least-squares solution of equation~(\ref{eq:lin}) with respect to $\mathbf{v}$
\begin{align}
    \mathbf{v}^* &= \arg\min_{\mathbf{v}}\|\mathbf{Ev}-\mathbf{m}\|_2^2\nonumber + \frac{\lambda}{2}\|v\|_2^2\\
    &= (\mathbf{E}^H\mathbf{E} + \lambda \mathbf{I})^{-1}\mathbf{E}^H\mathbf{m}
    \label{eq:l2}
\end{align}
with $^H$ denoting the Hermitian transpose. As the inverse of $\mathbf{E}^H\mathbf{E}$ is computationally demanding, the problem is typically solved in an iterative fashion. Optionally, the conditioning of the matrix inversion can be improved by addition of a small constant value $\lambda\geq0$ to the diagonal $\mathbf{E}^H\mathbf{E} + \lambda \mathbf{I}$, with $\mathbf{I}$ being the identity matrix. This type of modification is typically referred to as Tikhonov regularization~\citep{tikhonov1977}. For $\lambda=0$ the problem reduces to ordinary least squares. A numerically fast method to solve equation~(\ref{eq:l2}) is given by the conjugate gradient (CG) algorithm, outlined in Algorithm~\ref{alg:CG}. An full description of the CG algorithm can be found in~\citet{shewchuk1994}. The CG algorithm can be applied to problems of the form in equation~(\ref{eq:lin}) but requires a positive (semi-) definite matrix $\mathbf{E}$. This requirement can not be guaranteed for arbitrary encoding matrices $\mathbf{E}$. To this end, the CG algorithm is applied to the normal equation
\begin{equation}
    (\mathbf{E}^H\mathbf{E}+\lambda \mathbf{I})\mathbf{v} = \mathbf{E}^H\mathbf{m},
    \label{eq:normal}
\end{equation}
which yields the last-squares solution defined by equation~(\ref{eq:l2}).
Another advantage of the normal equation is that it has a positive (semi-) definite operator ($\mathbf{E}^H\mathbf{E}+\lambda \mathbf{I}$) by definition. Thus, the requirements for the CG algorithm are met.

\begin{algorithm}
\begin{algorithmic}[1]
\Initialize{
$\mathbf{v}_0 := 0$; \hfill $i := 0$; \hfill $i_{max} > 0$ \hfill $\mathbf{A} := \mathbf{E}^H\mathbf{E}$; \hfill $\mathbf{b} := \mathbf{E}^H\mathbf{m}$ \hfill $\mathbf{r}_0 := \mathbf{b} - \mathbf{Av}_0$; \hfill $\mathbf{p}_0 := \mathbf{r}_0$; \hfill $\delta=0$; \hfill $\epsilon > 0$}
\While{$i \leq i_{max}$}
\State $\delta \Leftarrow \frac{\mathbf{r}_i^H\mathbf{r}_i}{\mathbf{r}_0^H\mathbf{r}_0}$
\If{$\delta < \epsilon$ }
  \State break loop and return $\mathbf{v}_i$
\EndIf
  \State $\alpha_i \Leftarrow \frac{\mathbf{r}_i^H\mathbf{r}_i}{\mathbf{p}_i^H\mathbf{Ap}_i}$
  \State $\mathbf{v}_{i+1} \Leftarrow \mathbf{v}_i + \alpha_i \mathbf{p}_i$
  \State $\mathbf{r}_{i+1} \Leftarrow \mathbf{r}_i - \alpha_i\mathbf{Ap}_i$
  \State $\beta_i \Leftarrow \frac{\mathbf{r}_{i+1}^H\mathbf{r}_{i+1}}{\mathbf{r}_i^H\mathbf{r}_i}$
  \State $\mathbf{p}_{i+1} \Leftarrow \mathbf{r}_{i+1} + \beta_i\mathbf{p}_i$
  \State $k \Leftarrow i+1$
\EndWhile
\end{algorithmic}
\caption{The Conjugate Gradient algorithm.}
\label{alg:CG}
\end{algorithm}

If the noise correlation between receive coil channels can be estimated, e.g. from a separate noise scan, the coils and data can be pre-whitened to account for the correlation between different channels. This process creates virtual coils which can be used in the CG algorithm instead of physical coils without
requiring any other modifications~\citep{pruessmann2001}.

The conditioning of the problem and, thus, the convergence speed of the algorithm can be improved by
including a density compensation function into the reconstruction pipeline. This accounts for the typically non-uniform density in the k-space center of non-Cartesian sampling strategies. The diagonal density compensation matrix $\mathbf{D}$ can be included in the encoding matrix $\mathbf{E}$ by
\begin{equation}
    \bar{\mathbf{E}} = \mathbf{D}^{\frac{1}{2}}\mathbf{E},
\end{equation}
weighting each k-space signal by its spatial density. It should be noted that this introduces
a weighting in the data consistency, which then deviates from the noise optimal least squares
formulation. Intensity correction of the coil sensitivity profiles $\mathbf{I}$ can be included in analogy by
\begin{equation}
    \widetilde{\mathbf{E}} =\bar{\mathbf{E}}\mathbf{I}.
\end{equation}
Substituting $\mathbf{E}$ with $\widetilde{\mathbf{E}}$ in equation~(\ref{eq:normal}) gives the density and intensity corrected image reconstruction problem.
After convergence it remains to apply the intensity correction $\mathbf{I}$ to $\mathbf{I}^{-1}\mathbf{v}$ to obtain the final reconstruction result $\mathbf{v}$, which reduces to a point-wise division in image space. 

\subsection{Non-Uniform Fast Fourier Transform (NUFFT)}
If measured k-space points are acquired on a non-Cartesian grid, modifications to the standard FFT are necessary. The main steps involve:
\begin{itemize}
    \item Density compensation (optional).
    \item Gridding the non-Cartesian k-space to a regular but oversampled grid. Usually done with one of the following approaches.
    \begin{itemize}
        \item Convolution with a pre-computed kernel. Most common are Kaiser-Bessel based kernels~\citep{beatty2005}.
        \item Min-Max interpolation following ~\citet{fessler2003}
    \end{itemize}
    \item Standard FFT of the now Cartesian data.
    \item Deapodization - Accounting for intensity variations due to the convolution with the gridding kernel.
    \item Cropping to the desired Field-of-View (FOV).
\end{itemize}
These steps are generally referred to as non-uniform FFT (NUFFT). Even though it achieves the same computational complexity ($N\log{N}$) as the standard FFT, the computation is typically slower and scaling with dimensionality is worse.

\subsection{Data}
The evaluation in this work was performed on two different data sets. First, the algorithm was evaluated using radially sampled data provided by the organizers of the 2019 RRSG challenge. Second, during follow up work after the conclusion of the challenge, radial and spiral data including noise reference scans were acquired. These data sets closely follow the sampling trajectories and noise treatment of the original CG-SENSE paper, and were reconstructed with the consolidated reference implementations to evaluate their correctness and properties.

\paragraph{Challenge data}
The challenge data consist of two radial $k$-space data sets, one brain and one cardiac measurement, supplied in the \texttt{.h5} data format~\citep{hdf5}. The data set entries are ordered using the BART toolbox convention~\cite{uecker2015a}, i.e. for the data the dimensions [1, Readout, Spokes, Channels] and for the trajectory [3, Readout, Spokes], where the first entry encodes the three spatial dimensions. The distance between sampling points is $1/\text{FOV}_{os}$ and the entries run from $-N/2$ to $N/2$ with $N$ being the matrix size of the desired FOV. FOV$_{os}$ is the readout-oversampled FOV. The brain data consisting of 96 radial projections. 2D radial spin echo measurements of the human brain were performed with a clinical 3 T scanner (Siemens Magnetom Trio, Erlangen, Germany) using a receive only 12 channel head coil. Sequence parameters were: TR=2500 ms, TE=50 ms, matrix size = 256x256, slice thickness 2 mm, in plane resolution 0.78$\times$0.78 mm$^2$. FOV was increased to a matrix size of 300x300 after acquisition. The sampling direction of every second spoke was reversed to reduce artifacts from off-resonances. The cardiac data set consists of 55 radial projections acquired with a 34-channel coil on a 3 T system (Skyra, Siemens Healthcare, Erlangen, Germany). A real-time radial FLASH sequence with TR/TE = 2.22/1.32 ms, slice thickness 6 mm, 5 $\times$ 15 radial spokes per frame,  1.6 $\times$ 1.6 mm$^2$ resolution and a flip angle of 10$^\circ$ was used. Matrix size was set to 160 $\times$ 160 with 2-fold oversampling and a FOV of 256 $\times$ 256 mm$^2$, which was up-scaled by a factor of 1.5 after acquisition to fully contain the heart, leading to a reconstruction FOV of 384 $\times$ 384 mm$^2$ with a 240 $\times$ 240 matrix size.

\paragraph{Reference data}
In addition to the original challenge data two new data sets, a radially acquired heart data set and spirally acquired brain data set, were used in this work. The heart data was acquired at the Karolinska Institutet and the acquisition parameters are as follows:
Prototype bSSFP pulse sequence with golden-angle radial trajectory. Acquired at 1.5 T (Aera, Siemens Healthcare, Erlangen, Germany). Matrix size, $256\times256$ pixel, acquired pixel size 1.4 mm$^2$, 420 radial views, slice thickness 8 mm, TE/TR = 1.57/3.14 ms, flip angle $50^\circ$, receiver bandwidth 930 Hz/px. A 18-channel surface coil and a 12-channel spine coil (with 8 active elements) was used.

The brain data was acquired at ETH Zurich on a 3 T MR system (Philips Healthcare, Best, The Netherlands) using a 16-channel head coil with integrated magnetic field sensors (Skope MR Technologies and ETH, Zurich, Switzerland) with the following acquisition parameters: GE spiral trajectory with three interleaves, FOV = 22 cm, pixle size $1\times1$ mm$^2$ with $2$ mm slice thickness, TE/TR = 25/2000 ms, flip angle $90^\circ$. A total of 27121 samples per spiral were acquired.

The data sets are supplied as \texttt{.h5} files containing trajectories, multi-channel data, coil sensitivity maps, and noise covariance matrix. Written informed consent was obtained by all healthy volunteers following the local ethics committee's regulations. 

\subsection{Submissions}
\label{sec:submissions}
The following paragraphs will give a brief overview of the submissions with respect to the architecture and toolboxes used. Major differences between the submissions, especially implementation-wise, will be highlighted. The main features of each submission are summarized in Table~\ref{tab:algo}. 

To comply with the original algorithm, some sort of k-space filter function needs to be included in the reconstruction code. The most popular choice in all submissions is an arctan based filter function given by
\begin{equation}\label{eq:filt}
    f(k_x,k_y) = 0.5 + \frac{1}{\pi}\arctan{\frac{\beta*(k_c-\sqrt{k_x^2+k_y^2})}{k_c}}.
\end{equation}
If other filters are used, they are described in the corresponding paragraph of the submission.
The desired undersampling factor is attained by skipping every other acquired line for brain data to achieve factors of $\{1, 2, 3, 4\}$ compared to the acquired number of spokes. The heart data is undersampled by selecting the first $\{55, 33, 22, 11\}$ projections. Different realizations of undersampling for a given implementation are described in the corresponding paragraph.


\paragraph{University of California, Berkeley}
Based on Python this submission uses an interactive Jupyter notebook to run the reconstruction. 
The CG algorithm is implemented using the SigPy python package, which provides a high level abstraction interface to run code on CPUs or GPUs. Coil sensitivity profiles for the head scans are estimated via a root-sum-of-squares (SoS) approach, dividing each channel by the SoS reconstruction. The sensitivity profiles for the heart data are estimated using the BART toolbox~\citep{uecker2015a}.
Density compensation is done by taking the norm of each coordinate in the trajectory, resulting in a simple ramp function. The NUFFT algorithm is based on Fessler's min-max interpolation~\citep{fessler2003}.
Data is multiplied with the square root of the density compensation to assure adjointness of the forward and backward operator in the CG algorithm. The CG algorithm is terminated after $\{3,6,15,50\}$ iterations. No k-space filtering is performed after the CG reconstruction. Brain and heart data are undersampled by skipping acquired lines according to the desired undersampling factor.

\paragraph{Berlin Ultrahigh Field Facility (B.U.F.F)}
This Matlab based submission uses the NUFFT from the BART toolbox in the self-written CG implementation. Coil sensitivity estimation is performed via a SoS approach, dividing each channel by the SoS reconstruction. Density correction is performed by taking the Euclidean norm of each trajectory position in 2D, normalized by the maximum value. Intensity correction is achieved by normalizing the sum over the complex coil sensitivity to yield one. The CG algorithm is run on the brain data for 150 iterations for no acceleration and for 100 iterations for the accelerated cases. For the heart data set 40 iterations are used for all cases. 
The filter function for k-space position $k_x, k_y$ is given in equation~\ref{eq:filt} and $k_c$ being half the cropped FOV size, $\beta = 100$ are used as parameters. The filter is applied to the cropped FOV image after the CG algorithm has terminated.

\paragraph{Eindhoven University of Technology}
This submission uses Python to achieve the desired reconstruction results. The main package used for reconstruction is PyNUFFT which also supplies the CG algorithm. The internal NUFFT algorithm is based on Fessler's min-max interpolation~\citep{fessler2003}. Each coil channel is reconstructed independently using PyNUFFT and the final image is formed via a SoS approach. No density correction or coil sensitivity estimation is performed prior to reconstruction. The CG algorithm is terminated after 11 iterations.
No k-space filtering is performed after reconstruction. Brain and heart data are undersampled by skipping acquired lines according to the desired undersampling factor. 

\paragraph{Swiss Federal Institute of Technology Zurich (ETH)}
This implementation is built on code which was presented at past educational sessions and image reconstruction workshops of the ESMRMB and does not utilize any toolboxes to provide a white box implementation with full details provided to the user. The sensitivity maps are calculated from the fully-sampled center of the radial k-space data using an SVD-based approach by \citet{walsh2000}. The gridding operator is devised as a sparse matrix with a Kaiser-Bessel kernel function to utilize the fast matrix-vector multiplications in Matlab. A noise covariance matrix of the receive channels can be included in the reconstruction if available. The iteration is not regularized (e.g. by Tikhonov regularization), but stopped manually after a number of iterations determined by visual assessment of the intermediate result. In the last step, a k-space filter is applied to suppress noise from k-space areas with no acquired data.

\paragraph{Karolinska Institutet (KI)}
This submission is written in Matlab, using the object oriented capabilities of the language. The aim is to make the code as readable as possible. Coil sensitivities are estimated by dividing each coil element by the SoS of all coil elements. The CG algorithm is implemented from scratch using the same variable names and formalism as the original paper~\citep{pruessmann2001}, to help readers relate the code to the paper.
A ``SENSE-operator'' is implemented, which performs the steps outlined in the algorithm description provided in the original paper. A linear ramp-filter is used for sampling density compensation, and the gridding and Fourier transform is performed using the NUFFT routine, with Kaiser-Bessel interpolation, from the Michigan Image Reconstruction Toolbox (MIRT)~\citep{fessler2003}. No additional filtering of the k-space data is performed.

\paragraph{New York University (NYU)}
This submission is a straightforward implementation of the method described in~\citet{pruessmann2001} using available tools in Matlab. The implementation builds on the NUFFT toolbox by Jeff Fessler~\citep{fessler2003} for the radial reconstruction and on the Matlab portion of the gpuNUFFT package by Andreas Schwarzl and Florian Knoll~\citep{knoll2014} for the conjugate gradient algorithm implementation [1]. The theoretical density compensation function (k/max(k)) was used and coil sensitivities were estimated by applying an adaptive reconstruction~\citep{walsh2000} of single coil images. The CG algorithm is run for 100 iterations and augmented using Tikhonov regularization with a regularization weight of 0.2. If the residual values are below $1e^{-6}$ the algorithm is terminated. No k-space filtering is applied after the reconstruction. 

\paragraph{Stanford University}
The submission is based on Python and implements gridding via a Cython module. Gridding is realized via linear interpolation of a precomputed Kaiser-Bessel kernel function. The visualization of the reconstruction steps is done in an interactive Jupyter Notebook. Density compensation is realized via a simple ramp function computed from the Euclidean norm of the 2D k-space grid positions. The input data is multiplied with a Hamming window with the parameter M amounting to the number of samples on each spoke. Coil estimation is performed via a SoS approach, by filtering the data with a narrow-width Gaussian kernel ($\sigma=5\%$ k-space width), applying the inverse NUFFT to the filtered data and dividing each channel by the SoS reconstruction. In contrast to the other submissions, FOV cropping is performed to center the brain in the reconstructed image instead of cropping symmetrical around the center. The CG algorithm is run for 8 iterations for brain and heart data. 
The filter function for the normalized k-space position $k_x, k_y$ is given in equation~\ref{eq:filt} and $k_c=1$, $\beta = 100$ are used as parameters. The filter is applied in every iteration of the CG algorithm.

\paragraph{Graz, University of Technology, Institute of Computer Graphics and Vision (TUG H.)}
The submission is based on Python and uses the gpuNUFFT~\citep{knoll2014}, which is written in C++/CUDA to accelerate the reconstruction operator. Wrapper scripts to run the gpuNUFFT in Python are provided. The gridding in the gpuNUFFT is realized via nearest neighbour interpolation of a pre-computed Kaiser-Bessel function, where the gridding kernel size is set to 3. Coil sensitivity maps are estimated using ESPIRiT~\citep{uecker2014} from the BART toolbox~\citep{uecker2015a}. Intensity correction is applied by dividing the coil sensitivity maps by its SoS reconstruction, resulting in unit-norm along the coil channels. Density compensation is applied using a ramp filter estimated from the provided trajectories. To achieve adjointness of the forward and adjoint reconstruction operator $E$ and $E^H$, the raw data is multiplied by the square-root of the density compensation function. For reconstruction, 100 iterations are performed in the CG algorithm which was extended with Tikhonov regularization ($\lambda=0.2$).

\paragraph{Graz, University of Technology, Institute of Medical Engineering (TUG M.)}
The submission is based on Python and makes use of OpenCL to accelerate the reconstruction process. The Code is packaged and supports installation with automated dependency detection. Coil sensitivities are estimated from all acquired spokes using the ESPIRiT algorithm~\citep{uecker2014} implemented by the BART toolbox. Intensity correction is achieved by normalizing the sum over the complex coil sensitivity to yield one. As density compensation, a simple ramp function is used. The NUFFT is based on a Kaiser-Bessel gridding approach with linear interpolation between the pre-computed kernel points. The data is multiplied by the square root of the density compensation to ensure adjointness of the forward and backward application of the NUFFT. The CG algorithm is extended by an Tikhonov regularization and run for 300 iterations. The regularization weight $\lambda$ is chosen as 0.5.
The filter function for k-space position $k_x, k_y$ is given in equation~\ref{eq:filt} and $k_c = 25$,  $\beta = 100$ are used as parameters. The filter is applied to the cropped FOV image after the CG algorithm has terminated.

\paragraph{University of Southern California (USC)}
This Matlab implementation uses in-house written C/Mex function to implement the reconstruction algorithm. Gridding is realized via sinc interpolation combined with the Matlab FFT implementation. As density compensation, a simple ramp function is used. Coil sensitivity estimation is performed via a SoS approach using a 32-by-32 low resolution k-space center region and dividing each channel by the SoS reconstruction. Intensity correction is achieved by normalizing the sum over the complex coil sensitivity to yield one. The CG algorithm is run for $\{3, 6, 15, 50\}$ iterations for the brain data set and for 6 iterations for the heart data set.
The filter function for the normalized k-space position $k_x, k_y$ is given in equation~\ref{eq:filt} and $k_c=0.5$, $\beta = 100$ are used as parameters. The filter is applied to the cropped FOV image after the CG algorithm has terminated.

\paragraph{Utah Center for Advanced Imaging Research, University of Utah (Utah)}
This Matlab based submission uses a CPU/GPU accelerated NUFFT implementation, based on min-max interpolation of~\citet{fessler2003}. Coil sensitivities are estimated by the method of~\citet{walsh2000}. Sensitivity maps are normalized to give a sum of one along the coil dimension, accounting for intensity variations and ensuring adjointness. Gridding is performed via linear interpolation of a precomputed Kaiser-Bessel kernel. Density compensation amounts to a simple ramp. Reconstruction is run for $\{200, 100, 100, 100\}$ iterations for brain and 100 iterations for heart data. 
The filter function for k-space position $k_x, k_y$ is given in equation~\ref{eq:filt}. $k_c$ is chosen to be half of the cropped FOV and $\beta = 100$ is used as parameter. The filter is applied to the cropped FOV image after the CG algorithm has terminated.

\paragraph{Massachusetts General Hospital (MGH)}
This submission demonstrates how to use Matlab and modify the BART toolbox to perform a CG-SENSE reconstruction. Coil sensitivities are estimated using the ESPIRiT algorithm~\citep{uecker2014} of BART. In addition to the forward/backward application of the NUFFT, a faster Toeplitz-embedding based implementation is described. Reconstruction is performed by a modification of BART's "pics" method. Performance is demonstrated using all acquired projections of the brain data set only. No scripts to produce the desired images from undersampled data or for the heart data were submitted. 

\paragraph{Revised Submissions}
To avoid registration of individual submissions and eliminate errors due to necessary interpolation, participants were given the opportunity to submit revised code to account for differences in FOV and/or resolution between the reference and their submissions. Both, the original and the revised submission, are subsequently compared to the reference to show initial deviations and corrected images.

\subsection{Consolidated Implementation}
\label{sec:refimplement}
Accounting for the two major programming languages used throughout the submissions, reference implementations are developed both in Python and Matlab.

\paragraph{General steps}
To facilitate comparability between the two implementations, coil sensitivity profiles are pre-computed using the \citet{walsh2000} algorithm and all available projections. Density compensation is derived from the trajectory by gridding a k-space of ones followed by division of the maximum value. Taking the inverse of the gridded k-space yields the estimated density compensation function~\citep{jackson1991}. 
Reconstruction FOV and oversampling ratio are directly determined from the supplied trajectory. No scaling of the trajectory to specific intervals (e.g. $[-0.5, 0.5]$) is required. The apodization function is derived by Fourier transformation of the precomputed gridding kernel followed by normalization with the maximum value.
Furthermore, each iteration comprised intensity correction in image space based on the L2-norm of the sensitivity maps. In case of acquired noise reference data, noise pre-whitening is performed as a pre-processing step as described in~\citet{pruessmann2001}. The algorithm is initialized with an image of all zeros. Similar to the original work, the CG algorithm is terminated after a fixed number of iterations is reached which is chosen as 10 for all combinations of undersampling. As a final post-processing step, a k-space filter is applied after the last CG iteration to mask out ill-conditioned k-space areas, i.e. areas outside the circular support of the acquired data are masked out via hard thresholding. A supplementary step-by-step guide explaining details involved in each step of the reconstruction is provided online via a Jupyter notebook.

\paragraph{Matlab specific}
Sensitivity maps are assumed to be precomputed and read in at the start of the reconstruction. As in the original ETH submission, gridding is performed by a matrix-vector multiplication with a sparse matrix to reduce computation times for this time-critical operation, performed twice per iteration. The gridding kernel is based on a Kaiser window with width of 5 and 10000 pre-computed points. The value of the kernel for gridding a specific measurement point is determined via nearest neighbour interpolation. Furthermore, each iteration comprised intensity correction in image space as well as density correction in k-space, as described in the previous section. Explicit Tikhonov regularization is not included in accordance to the original paper.

\paragraph{Python specific}
If no coil sensitivity maps are supplied in the data file, receive coil sensitivities are estimated via the SoS approach, dividing each gridded coil image by the SoS reconstruction. To account for the typical smooth sensitivity profiles, the raw data is multiplied with a Hanning window with window width of 50 pixels, masking out high frequency components of the acquired k-space data prior to SoS reconstruction and coil sensitivity estimation. Optionally, the non-linear inversion (NLINV) algorithm~\citep{uecker2008} can be used estimate coil sensitivities prior to reconstruction. The pre-computed gridding kernel is derived using a Kaiser-Bessel function~\citep{beatty2005,jackson1991}. The kernel width is set to 5 and 10000 points of the window are pre-computed. The value of the kernel for gridding a specific measurement point is determined via linear interpolation of the pre-computed values. Optional Tikhonov regularization and a termination criterion can be enabled by setting the corresponding values larger than zero in the configuration file. 

\subsection{Numerical Comparison}
The results from each submission were compared on a pixel-by-pixel basis to the Python consolidated reference implementation. To account for possible intensity variations and outliers, each image was normalized by its 0.95 quantile intensity value prior to the difference operation. Reconstructions not matching the FOV of the reference were cropped prior to the difference operation. Cropping was performed symmetrically around the image center.

The two reference implementations were compared to each other in a similar fashion. Additionally, the structural similarity index measure (SSIM), using the parameters as suggested by~\citet{wang2004}, and normalized root-mean-squared error (NRMSE), defined in equation~\ref{eq:NRMSE}, are used as metrics to compare the two implementations.

\begin{equation}
    \label{eq:NRMSE}
    \text{NRMSE} = \frac{\sqrt{\frac{1}{n^2}\sum_i\left(|x_i|-|x^{ref}_i|\right)^2}}{\frac{1}{n^2}\sum_i |x^{ref}_i|}
\end{equation}

\subsection{Code and Data Availability Statement}
All data and code used in this paper are available online. Availability and associated Digital Object Identifiers (DOIs), if applicable, are listed below:
\begin{itemize}
\item Challenge submissions~\url{https://ismrm.github.io/rrsg/challenge_one/}. 
\item Reference implementations~\url{https://github.com/ISMRM/rrsg_challenge_01}
\item Original data~\url{https://ismrm.github.io/rrsg/challenge_one/}.
\item Supplementary spiral brain and radial cardiac data~\url{Zenodo.URL}.
\item Evaluation scripts to produce the figures~\url{Zenodo.URL}
\item All code and data is licensed under the MIT License (\url{https://en.wikipedia.org/wiki/MIT_License})
\end{itemize}

\section{Results}
\subsection{Submissions}
\begin{figure}
    \centering
    \includegraphics[width=0.7\textwidth]{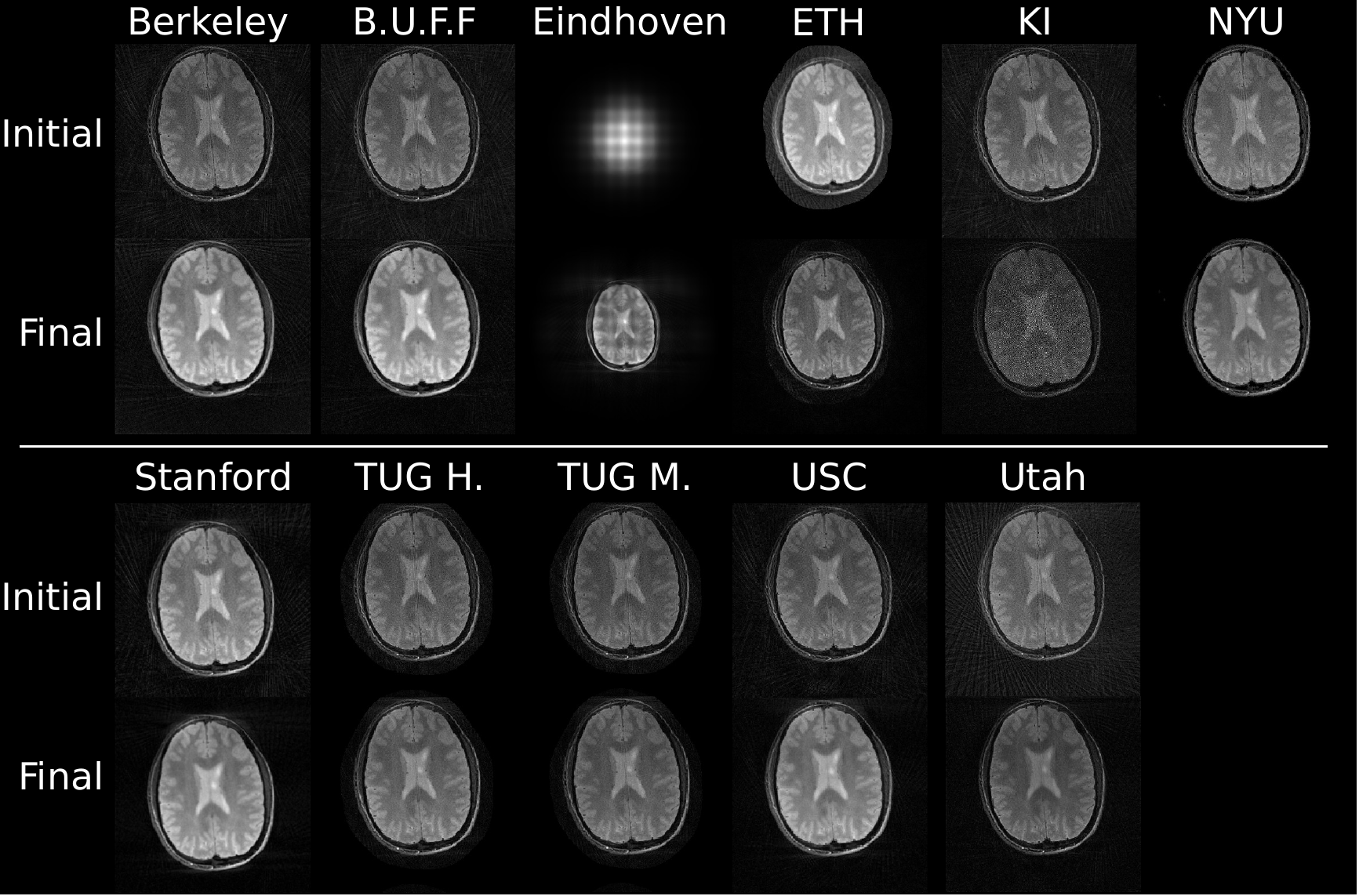}
    \caption{Example images after evaluation of the right-hand-side of 
equation~\ref{eq:normal} (Initial) and after termination of the algorithm for 
(Final) for each submission. Shown are results for acceleration factor of 2 of 
the supplied challenge brain data. All results are shown as they are returned by 
each algorithm. Visually observable differences include intensity variations 
among the reconstructions as well as some image center shifts and FOV 
differences. In addition, some reconstructions utilized image masks for the 
background.}
    \label{fig:brainrecosub}
\end{figure}

\begin{figure}
    \centering
    \includegraphics[width=0.7\textwidth]{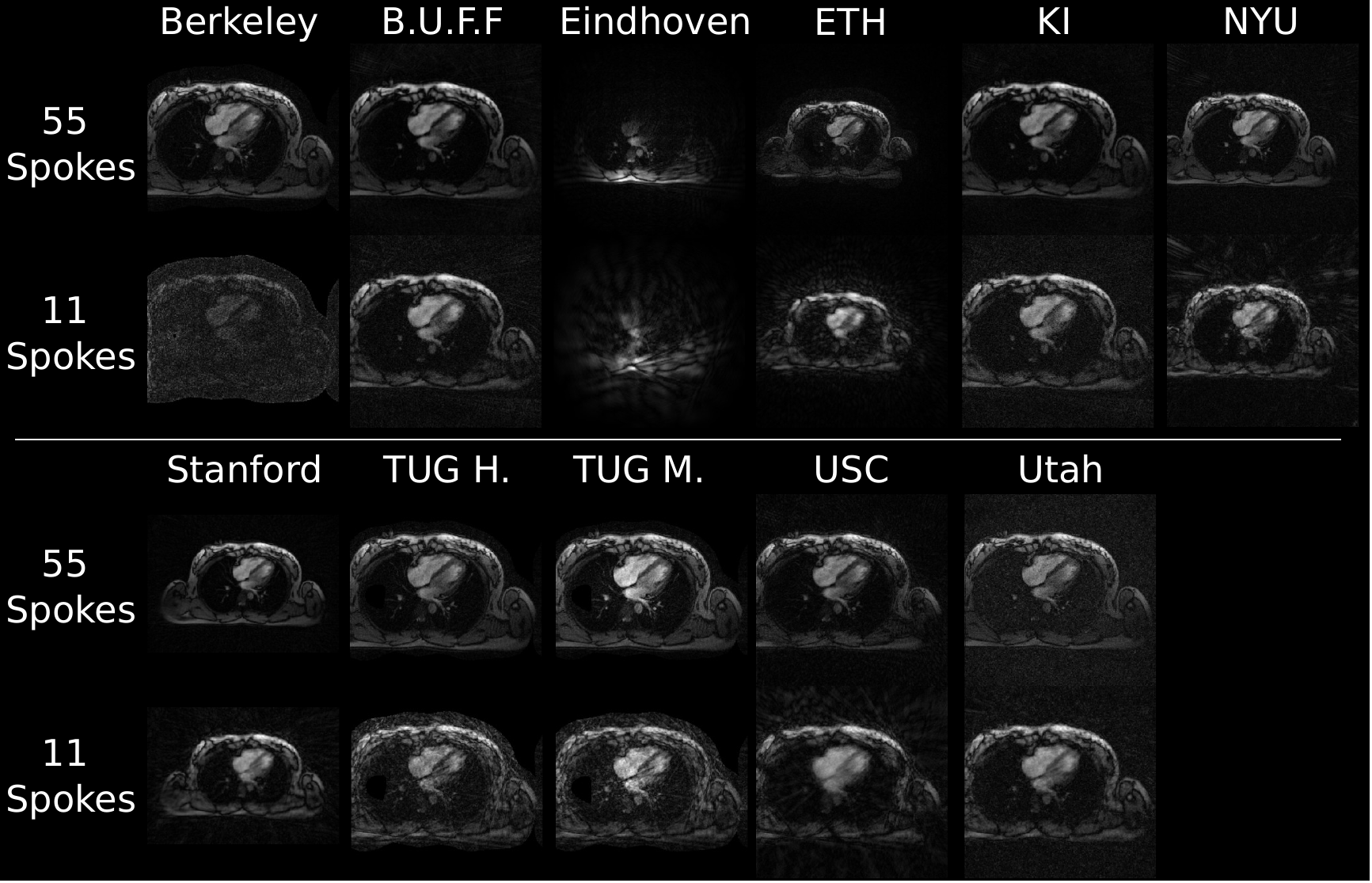}
    \caption{Example reconstruction results for each challenge submission. Shown 
are results using 55 and 11 spokes of the supplied challenge heart data. 
Visually observable differences amount to FOV changes as well as image center 
changes. Intensity variations are not as severe as in the case of brain data. 
Again, some reconstructions made use of a mask to suppress background signal.}
    \label{fig:heartrecosub}
\end{figure}

Example reconstruction results for an acceleration factor of 2, showing the image after evaluation of the right-hand-side of equation~\ref{eq:normal} (Initial) and after convergence of the algorithm (Final), from each submission are given in Figure~\ref{fig:brainrecosub} and Figure~\ref{fig:heartrecosub} for brain and heart data, respectively. All results are displayed with a window width from the minimum to the maximum occurring value in each image. Visually, intensity variations are noticeable owing to the different maximum values, however, contrast between different tissue seems to be similar in all submissions. Some submissions also use different FOVs for the brain (Eindhoven - no cropping, ETH - cropped to 340x340, Stanford - cropped asymmetrical to 300x300 ) compared to the others or different matrix sizes in the same FOV (USC - 256x256 and Utah - 512x512). No major structural differences are observable in the reconstruction except for the case of Eindhoven. The brain reconstruction from KI did neither use Tikhonov regularization nor early stopping, and the k-space was not filtered, resulting in a noisy appearance compared to other submissions. In addition, it shows a slight rotation to the left. 

\begin{figure}
    \centering
    \includegraphics[width=0.85\textwidth]{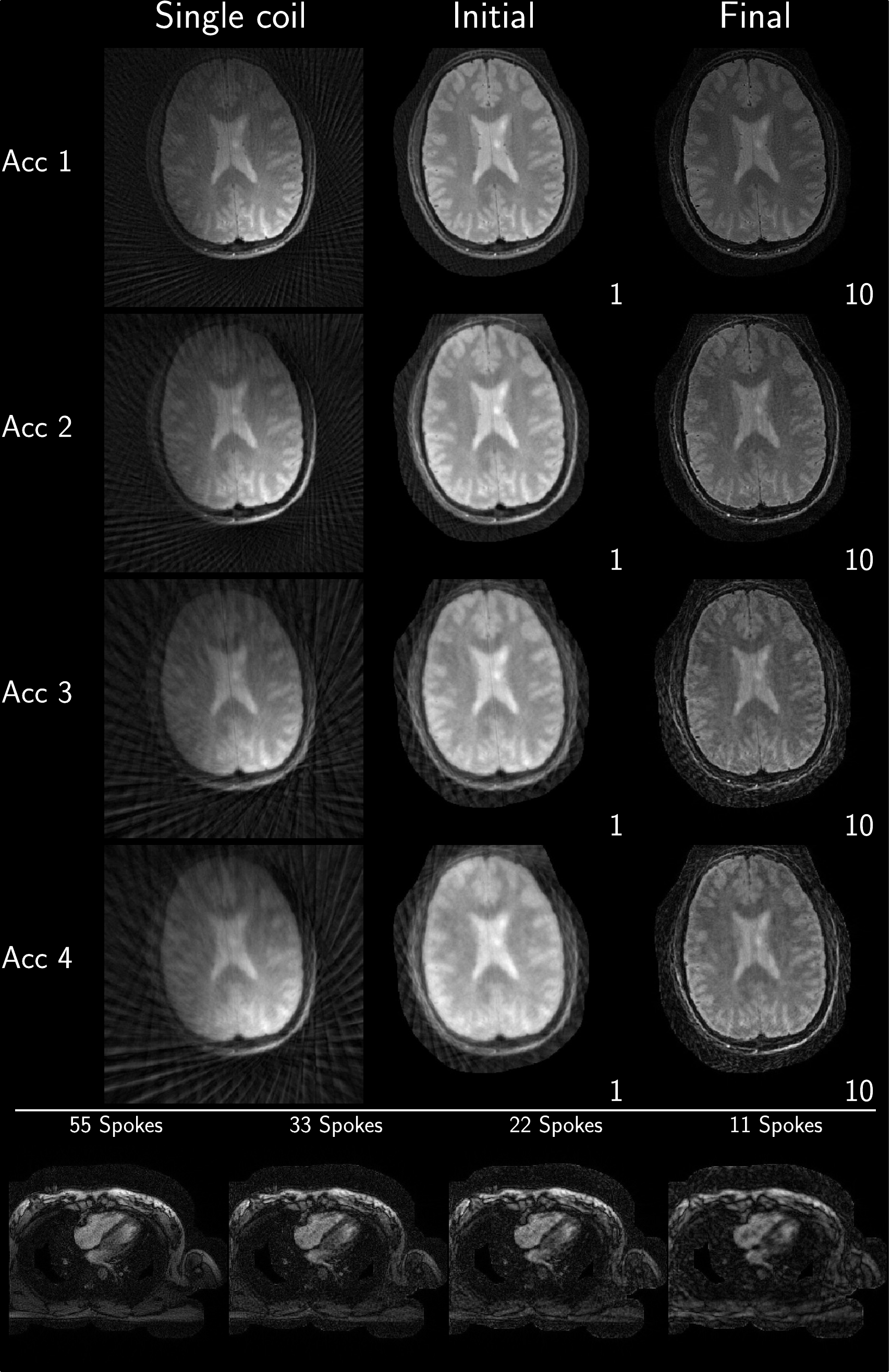}
    \caption{Consolidated reconstruction results using the Python 
implementation.}
    \label{fig:refpython}
\end{figure}

\begin{figure}
    \centering
    \includegraphics[width=0.85\textwidth]{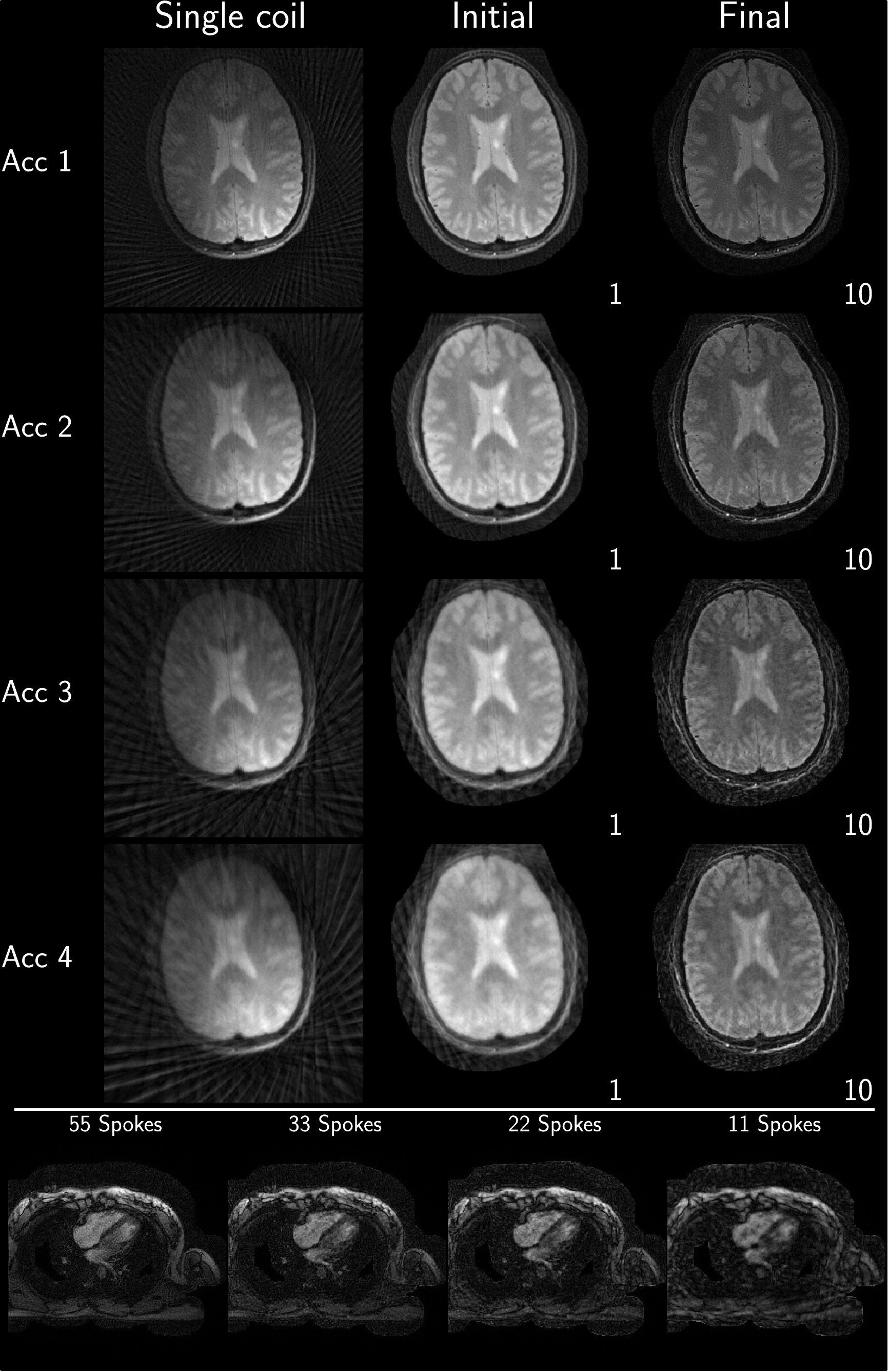}
    \caption{Consolidated reconstruction results using the Matlab 
implementation.}
    \label{fig:refmatlab}
\end{figure}

For the heart data, more differences are observed. Firstly, FOV differences occur more frequently (Eindhoven - cropped to 320x320, ETH - cropped to 360x360, NYU - cropped to 300x300, Stanford - asymmetric crop to 240x336). Secondly, matrix size in the same FOV and thus resolution is changed by some submissions (Berkeley - 300x300, USC - 256x256, Utah - 320x320). The heart reconstruction results for 11 spokes from Berkeley seem to be more noisy than the others. The reconstructions using the reference implementations, are given in Figure~\ref{fig:refpython} and Figure~\ref{fig:refmatlab} for Python and Matlab, respectively. Neither intensity nor contrast variations are visible between the two reference implementations. 


\subsection{Differences to Consolidated Implementation}
 Visually, no major differences to the submissions are visible. Pixel-wise difference plots are shown in Figure~\ref{fig:diffBrain} and Figure~\ref{fig:diffHeart}. Reconstructions from Eindhoven, KI, USC, Stanford and Utah show some misalignment caused by image center shift, rotation or matrix size differences compared to the reference after cropping to the desired resolution of $300{\times}300$ pixels. Small intensity variations across the brain are visible in the final step of the TUG M. reconstruction. Reconstructions from Berkeley, B.U.F.F., ETH, NYU and TUG H. show the least deviations to the reference. Visually no difference in the brain tissue can be seen. Initial steps seem to show good agreement if image alignment matches with only minor intensity difference in some submissions. Heart data shows overall more differences, especially in areas of low SNR. The heart itself seems consistent between most reconstructions. At the highest acceleration, differences become more pronounced.  

\begin{figure}
    \centering
    \includegraphics[width=0.9\textwidth]{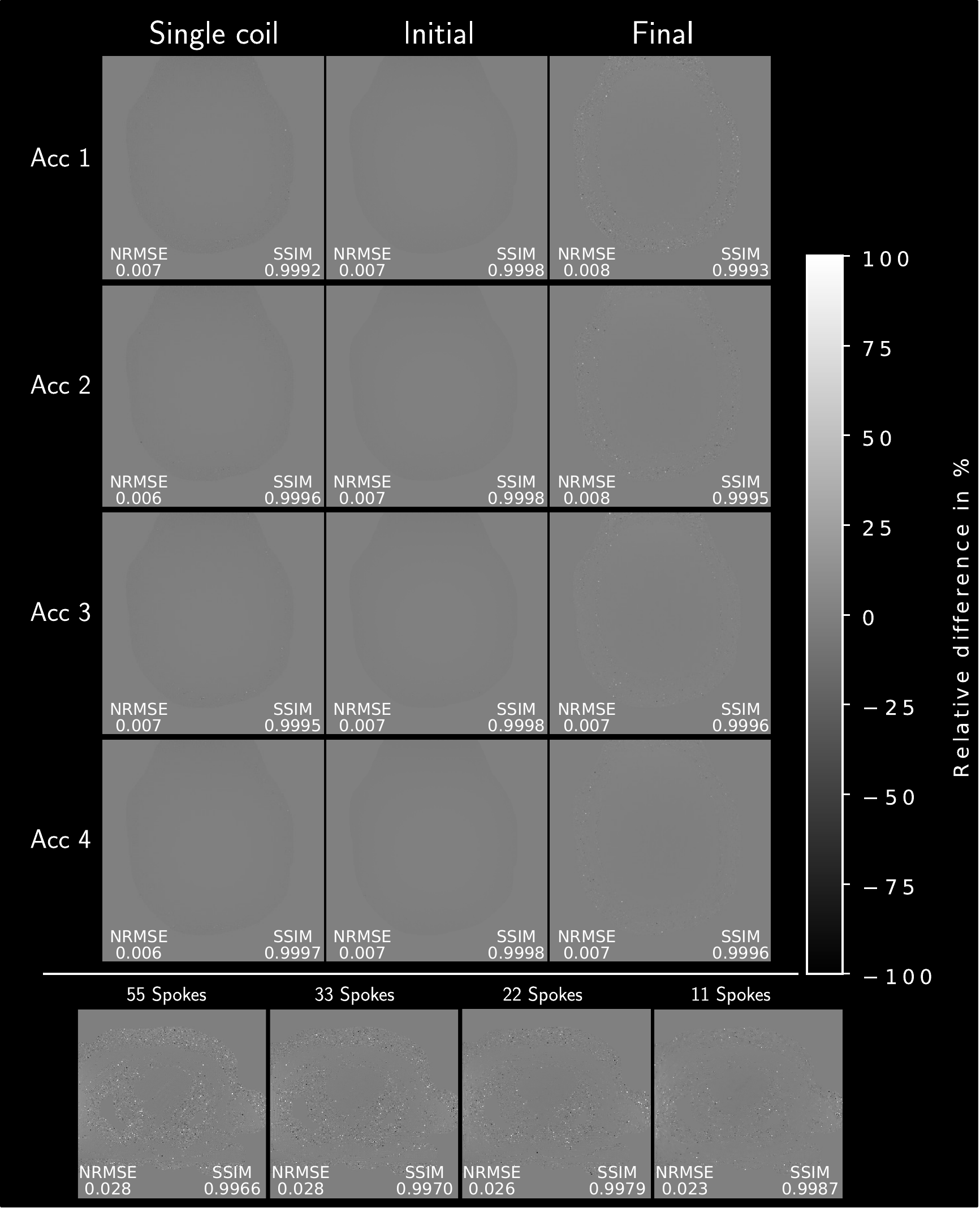}
    \caption{Relative pixel-wise difference between the Matlab reference 
reconstruction results and the Python reference implementation. To account for 
different intensities, all images were normalized prior to the difference 
operation. SSIM and NRMSE values between the two references are given next to 
each image. Metrics are computed from values within a binary mask, containing 
the brain and heart, respectively.}
    \label{fig:diffRef}
\end{figure}

The pixel-wise comparison between the two reference implementations in Figure~\ref{fig:diffRef} for brain (top part) and heart (bottom part) data shows the overall excellent accordance between the Matlab and Python reconstruction results. No major differences are visible in any of the images. The single coil images and initial images show very slight intensity differences. The visual impression is supported by high SSIM values (0.9987-0.9998) and small NRMSE values (0.006-0.028). The highest differences are visible outside of the brain at the border of the used image mask. Similar excellent accordance between both reconstructions is achieved for heart data. NRMSE and SSIM are comparable to the brain data but areas with little to no signal outside the body show slight, noise-like deviations.

\begin{figure}
    \centering
    \includegraphics[width=0.8\textwidth]{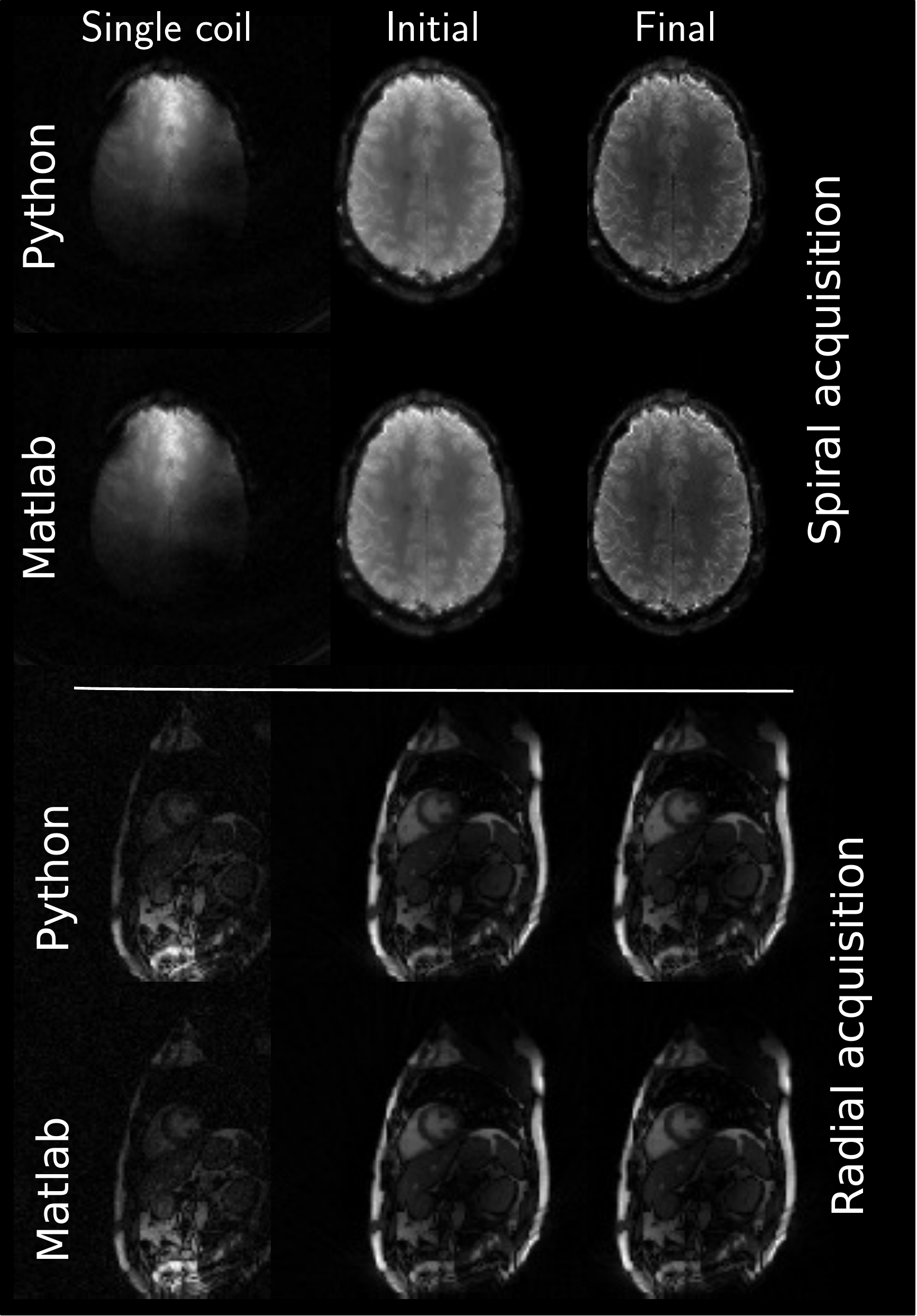}
    \caption{Example reconstruction results of the two additional supplied data 
sets. Top rows show the spirally acquired brain data set and bottom rows show 
radially acquired cardiac data. Both reconstructions included noise 
pre-whitening prior to reconstruction from a dedicated noise scan preceding 
image acquisition. Windowing is performed between the minimum and maximum 
intensity value in each image.}
    \label{fig:newdata}
\end{figure}

Finally, Figure~\ref{fig:newdata} shows the two additionally supplied data sets. Both reference methods are able to produce clean images and visually, no differences can be observed.


\section{Discussion}
\subsection{Achievements of the Challenge}
The first ISMRM reproducibility challenge led to twelve submissions from research groups spread across many countries. In addition to the paper challenge a questionnaire was opened \footnote{https://blog.ismrm.org/2019/04/15/reproducible-research-study-group-questionnaire/}, regarding reproducible research, which showed that the majority of the 71 participants (77.5\%) sees a reproducibility problem in their research area. This proves that scientists are aware of the problem of reproducibility of research and how hard it can be to recreate paper results without access to code or data. Furthermore, the challenge led to the production of two consolidated implementations of the seminal paper, written in the two most commonly used programming languages across all submissions. However, it also raised the question of what concretely \emph{reproducibility} means and how to measure similarity between different images. Even though many different toolboxes and reconstruction approaches were used by the participants the visual appearance of the reconstructions is very similar.

\subsection{Difference of Submissions}
\paragraph{Imaging Parameter}
As no desired FOV was given, some problems arose with the correct choice of the reconstruction FOV. It is common to assume an oversampling of 2 compared to the radial acquisition but for the supplied data, this was not the case. The brain was oversampled by a factor of $1.70\overline{6}$ and the heart by $1.\overline{3}$. These factors could be derived from the supplied trajectory but were not taken into account by all submissions. Larger FOVs can be easily corrected for by simply cropping to the desired FOV, which is commonly done symmetrically around the image center. All decided to crop in such a way, except for the submission of Stanford which crops symmetrically around the object. Although such an approach leads to a centered image, it can be tedious. The main reason for cropping the FOV is to crop away areas with aliasing, which typically folds back at the edges of the oversampled image grid~\cite{jackson1991, osullivan1985}. The aliasing gets worse with increased distance from the image center due to the amplification from the deapodization function and, depending on the gridding kernel, additional errors from outside of its pass-band can be introduced at the image edges~\citep{beatty2005,jackson1991, osullivan1985}. 

\begin{figure}
    \centering
    \includegraphics[width=\textwidth]{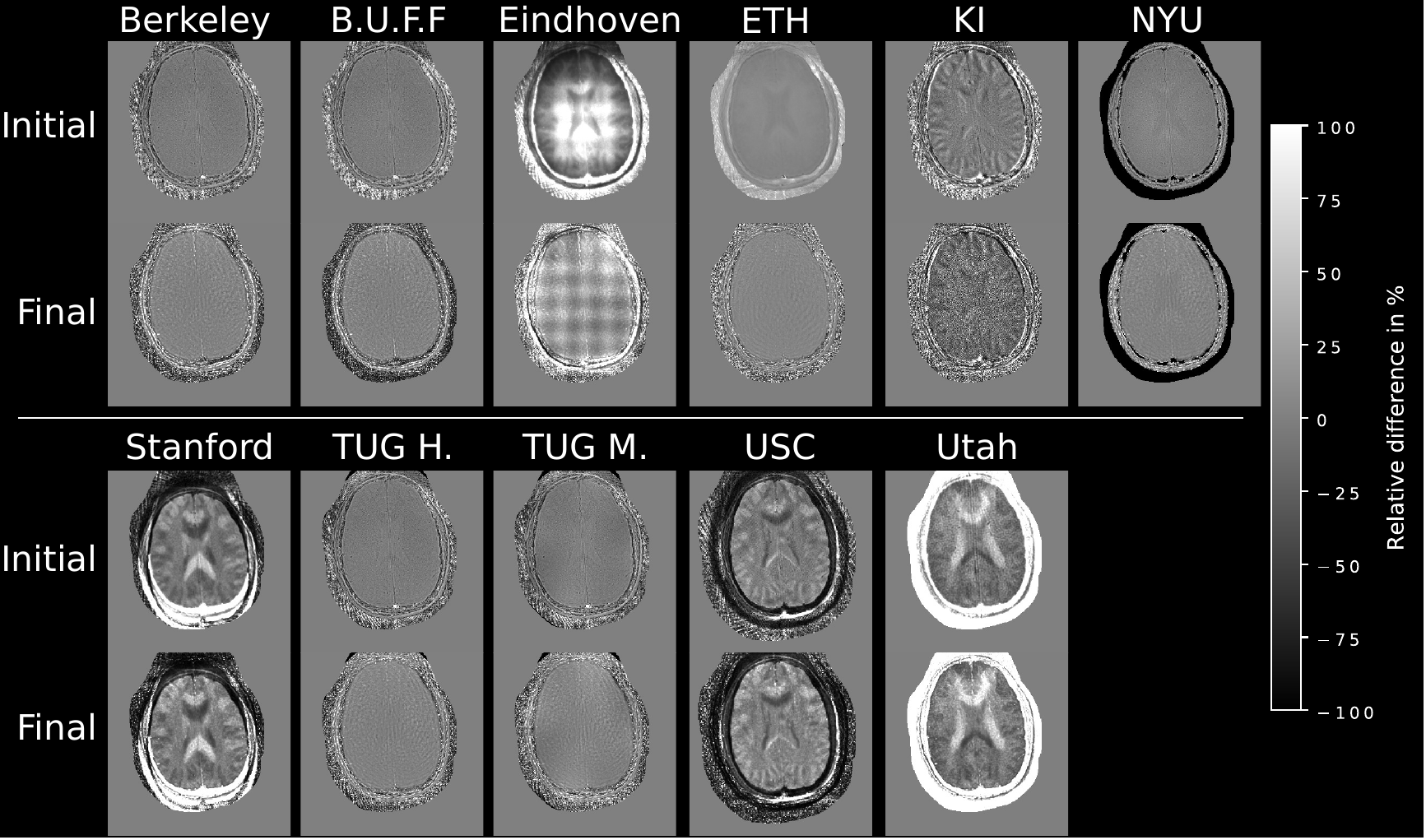}
    \caption{Relative pixel-wise difference of the example reconstruction 
results to the Python reference implementation. To account for different 
intensities, all images were normalized prior to the difference operation, 
however the submissions were not registered in terms of lateral shifts or 
rotation. Still, most reconstructions do not show substantial structural 
differences to the reference.}
    \label{fig:diffBrain}
\end{figure}

\begin{figure}
 \centering
 \includegraphics[width=\textwidth]{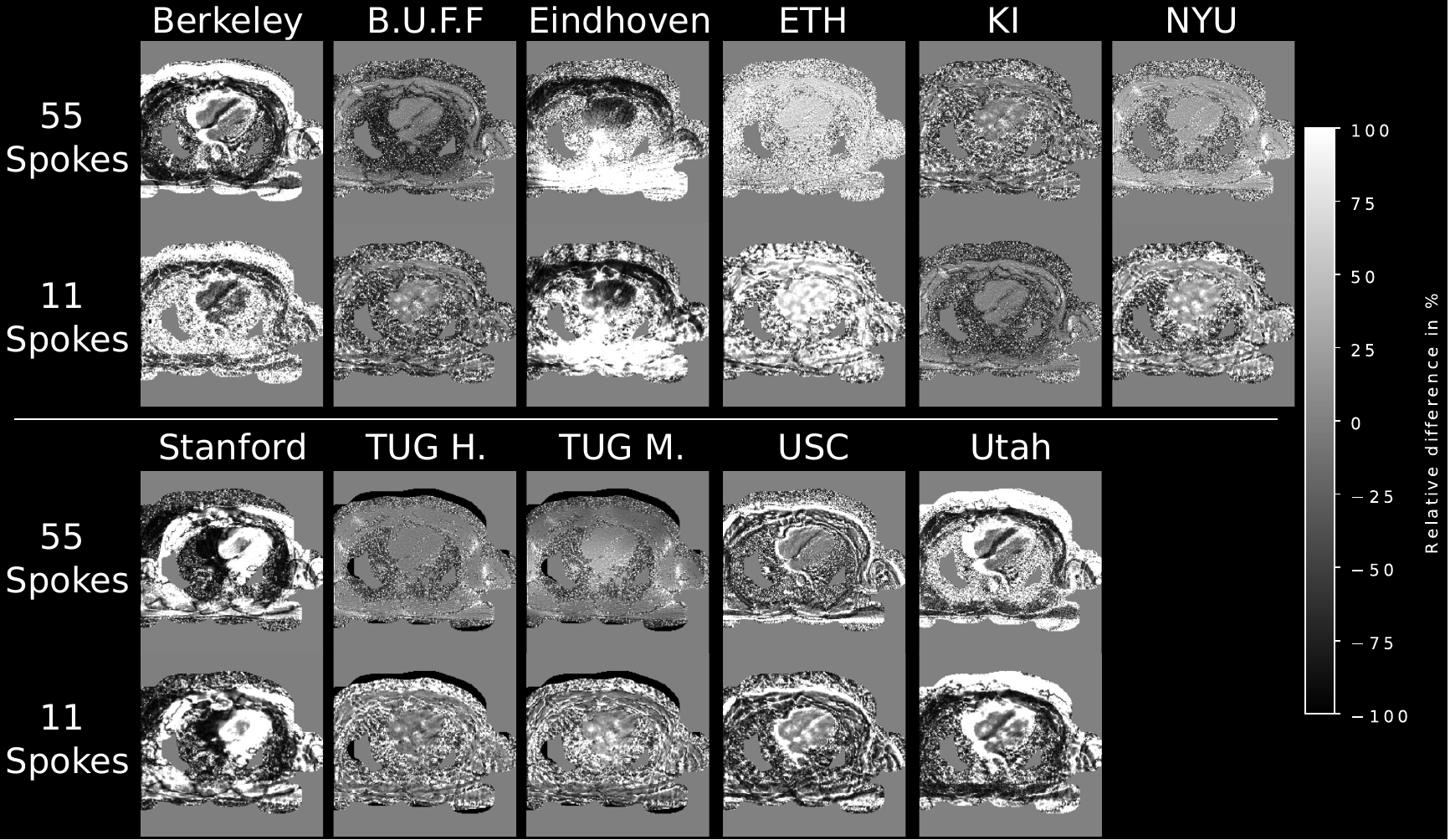}
  \caption{Relative pixel-wise difference of the example reconstruction results 
to the Python reference implementation. To account for different intensities, 
all images were normalized prior to the difference operation. Most 
reconstructions show similar structural information in the heart itself but 
differ in low signal areas. With increased acceleration differences become more 
pronounced.}
 \label{fig:diffHeart}
\end{figure}

\begin{figure}
    \centering
    \includegraphics[width=0.9\textwidth]{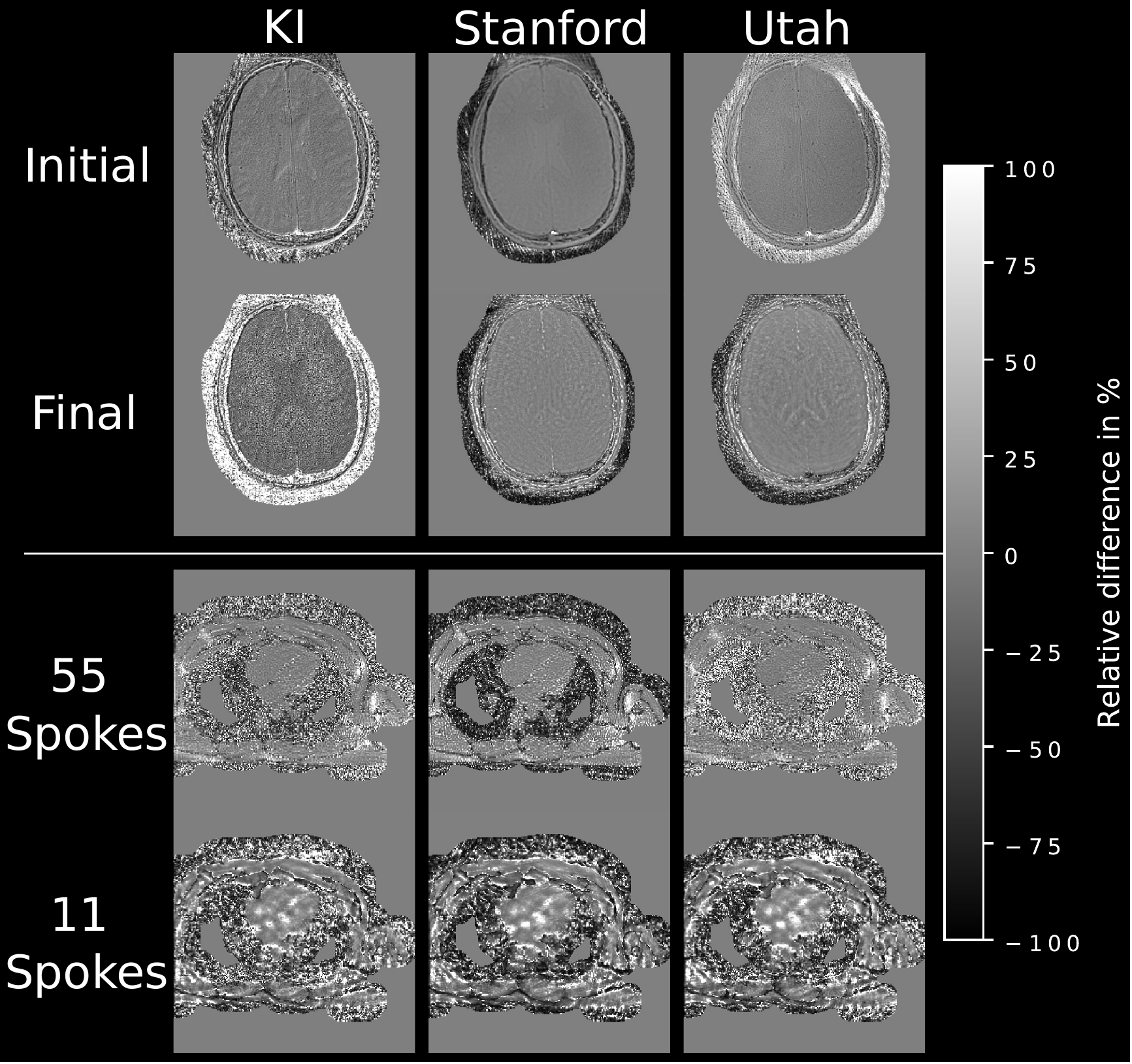}
    \caption{Relative pixel-wise difference of revised submissions, correcting FOV and/or trajectory related deviations to the reference implementation. In contrast to the initial submissions only minor deviations are visible.}
    \label{fig:diffRevised}
\end{figure}

\paragraph{Gridding/NUFFT}
As can be seen in the difference images in Figure~\ref{fig:diffBrain} and Figure~\ref{fig:diffHeart}, reconstructions with larger FOV show little or no structural differences to the references. A more concerning modification is the change of resolution as such changes can potentially lead to interpolation related changes in visual appearance of the image. A possible source of such an increased or decreased pixel size lies in the way, how acquired data points are placed in the k-space via the gridding operation. If an oversampled grid is defined but the location of the samples in the trajectory is not correctly altered to span the whole range of this oversampled k-space, only the central part will be filled. Similar, if the points-to-grid lie outside of the desired k-space, they either are not gridded at all or enter on the opposite side of k-space, depending on the used boundary conditions. This leads to an interpolation in image space and an artificially altered resolution, i.e. interpolation to higher or lower resolution, respectively.  Small structural differences in the submissions may stem from different treatment of the supplied k-space trajectory. Normalization of the k-space coordinates, as is done in many submission, might lead to modifications of the actual k-space position if done independently for each of the spatial dimensions. This can lead to a rotation or distortion of the reconstructed image. Such differences can not easily be corrected in the final images as those would involve some sort of interpolation to the desired matrix size or image registration, introducing errors related to the interpolation kernel. Therefore, no attempt was made to correct for different resolutions in the final image, leading to rather large deviations in the pixel-wise difference maps. Revised submissions in Figure~\ref{fig:diffRevised}, accounting for deviations in trajectory handling and/or FOV, show numerical differences for brain and heart data which are in-line with most of the other submissions. This suggests that the rather huge differences in the original submissions solely stem from wrongly treated trajectory information or FOV cropping.


\paragraph{Algorithmic}
The increased noise in the KI reconstruction may stem from the large amount of CG iterations combined with not using any regularization or k-space filtering. Running the CG algorithm for too many iterations leads to increased noise in the final reconstruction. This can also be seen in the heart reconstruction from 11 spokes from the Berkeley submission. Thus, early stopping is used as regularization in the original publication. Another form of regularization used in the submissions is plain Tikhonov regularization based on the $L_2$-norm of the image, i.e. $\lambda>0$ in equation~\ref{eq:normal}. The regularization parameter, typically termed $\lambda$, is used to balance between data costs and regularization. Even though this is not included in the original publication, results from submissions with Tikhonov regularization show only minor differences to the reference using early stopping, see Figure~\ref{fig:diffBrain} and Figure~\ref{fig:diffHeart}. However, choosing a correct regularization parameter can be a challenging task. A too large value for the regularization parameter can lead to slow convergence and may be the reason for residual intensity variations in the TUG M. submission for both, brain and heart data. On the other hand, choosing a regularization parameter is usually easier than choosing a number of iterations, because it can be done based on sound principles~\cite{golub1979,hansen1993,hansen1998}.

The solution of the inverse problem of finding an image from non-uniform acquired data highly depends on the quality of the pre-computed coil sensitivity profiles. A wrong or inaccurate estimate ultimately leads to poor reconstructed images. Visual effects include intensity variations or signal voids in the image. In addition, the estimated coil profiles influence the solution via the intensity normalization, directly estimated from the coil profiles. Thus, all provided data also contain estimated coil sensitivity profiles, which were used to generate the reference reconstruction results in both algorithms. 

\paragraph{Evaluation}
As images are typically given in arbitrary units, a direct numerical comparison can be challenging. As a result, image intensity normalization was applied. However, if normalization fails due to outliers or, in a more general sense, due to deviations with respect to the expected intensity histogram, it can lead to a false impression of rather large differences. This may be the reason for the increased deviations in the ETH submission of the heart data compared to the brain data as can be seen in the bright error map in Figure~\ref{fig:diffHeart}. To this end, no numerical metrics were used to compare submissions to the reference implementation as these would suffer even more from intensity variations or image shifts.

\subsection{Reference Implementations}
During the development of the reference implementations, we indentified that processing steps related to gridding yield the largest deviations, e.g. trajectory normalization, apodization correction, and gridding kernel normalization. The largest deviations were associated with the normalization of the trajectory to a specific range. The least deviations can be achieved without any modifications of the supplied trajectory, i.e. taking the k-space locations as is and adapting the gridding to account for the increased range of possible values (e.g. $>>|0.5|$).

The two reference implementations show no major difference inside the brain as evident in Fig.~\ref{fig:diffRef}. A very slight intensity difference for single coil and initial images can be seen which might be related to the apodization correction. Minor implementation details, such as the linear interpolation of the gridding kernel versus the nearest neighbor interpolation or the FFTW~\citep{frigo1998} in Matlab versus the FFTPACK based algorithm~\citet{cooley1965,bluestein1879} of Python, in combination with the iterative steps of the algorithm may lead to the remaining differences.
The heart data shows overall good agreement with increased deviations in areas with little to no signal, either inside the lungs or outside the body. The area of interest, the heart, shows no substantial difference between the two reference implementations. The SSIM and NRMSE metrics are computed within the same binary mask used for reconstruction. Thus, even better values for these metrics are expected if only the brain tissue itself is evaluated. The same is true for the heart reconstruction. Cropping the area to only include reliable pixels, i.e. pixels with high enough signal, SSIM values could be further improved.

To this end, the implementations of the CG-SENSE algorithm in Matlab and Python can be considered equally accurate and thus the submitted algorithms were compared to just one of the two references, the Python based implementation.

\subsection{Licensing Code and Data}
When the challenge was initiated, very little constraints were implied on how data could be used and code should be provided, to enable wide-spread participation. However, in retrospect, a wide-spread adoption and re-use of the data and code submissions created by the challenge requires some consideration of licensing, in order to stand on firm legal footing.

This is because if no license is specified, the owners of code or data retain all copyright, and have to give explicit permission for its use. But in the context of reproducibility, making software open-source and re-usable for other researchers is key. Two classes of software licenses are best suited for this cause: copyleft licenses, such as the General Public License (GPLv3), or permissive licenses, such as the MIT license.

There are good resources explaining the differences between those~\citep{sonnenburg_ossneed_2007}, including a very accessible website how to choose one: \url{https://choosealicense.com/}. In brief, MIT has the least restrictions and simplifies commercial use, whereas GPL puts emphasis on keeping code open-source, i.e., if one builds on GPL-licensed code, one has to make it publicly available, even in commercial settings. This also means that MIT-licensed code can be used within a GPL-project, but not the other way around, and one might have to choose GPL as a license then.

For sharing data, the situation is complicated by the fact that data might be considered part of software and documentation, or work of creative art, for which the class of creative common (CC) licenses were envisaged (\url{https://creativecommons.org/choose/}). If the source should be attributed and any use granted, including alterations and commercialization, CC-BY 4.0 is an appropriate choice for data. Recently, the Open Data Commons~(\url{https://opendatacommons.org/}) initiative of the Open Knowledge Foundation started to provide specific licensing tools for data.

For this challenge, we decided to license all data and code of the reference implementation under the MIT license, in order to keep licensing as simple and permissible as possible. We list the choice of licensing for all contributions in table~\ref{tab:algo}.

\subsection{Future Impact}
The first RRSG challenge has already been met with a very positive response, both in reproducing the selected publication, but more importantly in bringing together a community of researchers who are interested in reproducible science and MR image reconstruction. On top of that, we also see very practical use of its outcome in the future, as a benchmark for novel implementations of CG-SENSE. The clear definition of the challenge and its outcome measures, combined with the resulting reference implementation and comparison code, might encourage researchers to put their own reconstruction tools to the test. It should be noted that proper tuning of iteration numbers or regularization parameters is indispensable if reference methods are applied to new data to enable a fair comparison.

In fact, researchers have already started to adopt this idea and created submissions after the official challenge had ended. A recent effort by the Hamburg University of Technology (\url{https://github.com/MagneticResonanceImaging/ISMRM_RRSG}) demonstrates reproducibility of the CG-SENSE algorithm in the modern programming language Julia~\citep{bezanson_julia_2017}  (\url{https://julialang.org/}) utilizing their MRIReco.jl reconstruction package~\citep{knopp_mpirecojl_2019}. We believe that this could become a general model for future software publications to use proposed example data and outcome measures of reproducibility challenges in order to show performance and scope of these tools in a more standardized fashion.

Reproducibility of image reconstruction in MRI can be challenging, especially with the increased complexity of the used algorithms. Even though the description in a paper allows to re-implement the reconstruction algorithm, a lot of details may be not stated explicitly and can lead to unexpected outcome, e.g. exact step-sizes used in optimization, scaling of gradients, internal SNR estimates and other pre- and post-processing steps. These problems arise in many iterative fitting strategies throughout the whole field of MRI research, e.g. quantifying tissue parameters, estimating perfusion/diffusion metrics, just to name a few. 

Quantifying tissue parameters, more specifically the T$_1$ relaxation constant, is also the aim of the "Reproducibility Challenge 2020" of the study group. Reproducing exact quantitative values at multiple sites is challenging due to small variations in measurement imaging hardware and software. The challenge aims to identify the sources of variation and tries to standardize T1 mapping across multiple sites.

\section{Conclusion}
The present work shows that reproducing research results without access to the original source code and data leaves room for interpretation. Even though visual differences are minor for most submissions, a lot of deviations in various implementation details can be observed. During the evaluation it became clear that the task of comparing the submissions to each other is by no means trivial. Seemingly minor details, such as maximum image intensity or FOV and resolution can lead to huge deviations in a pixel-wise comparison even though visually differences are small. This raises the question of what can be considered ground truth. A question, which has no clear answer if neither original code nor data are available. A consolidated implementation can be used as substitute in such cases, as done in the present work. From what we have learned in this first reproducibility challenge, our recommendation is publishing not only papers but also code and data to make sure research is really reproducible.

\section*{Author contributions}

OM wrote the manuscript.

AF and FP performed measurements for the consolidated reference implementations.

OM implemented the Python reference implementation.

FP implemented the Matlab reference implementation, LK reviewed the code. 

SH implemented a step-by-step guide in a Jupyter notebook based on the Python reference.

OM, SB, AF, KH, FP, LK, ML, YT and KW participated in the reproducibility challenge.

DG, MU and FK organized the challenge, provided the data and analyzed the results.

FK oversaw the design and implementation of the reference solutions of the challenge and the writing of the manuscript.

All authors proofread the manuscript.

\section*{Acknowledgements}
The authors would like to thank 
\begin{itemize}
    \item[Berkeley] {Miki Lustig, Ekin Karasan, Suma Anand, Volkert Roeloffs}
    \item[ETH] {Thomas Ulrich, Maria Engel, S. Johanna Vannesjo, Markus Weiger, David O. Brunner, Bertram J. Wilm, Klaas P. Pruessmann, as well as Felix Breuer and Brian Hargreaves, who contributed the sparse matrix gridding implementation}
    \item[MDC/B.U.F.F. - Berlin] Ludger Starke
    \item[MGH] Gilad Liberman
    \item[USC] Namgyun Lee
\end{itemize}
for participating in the first ISMRM Reproducible Research Study Group challenge and spending their valuable time to reproduce the CG-SENSE algorithm. 

This work was supported by the NCCR “Neural Plasticity and Repair” at ETH Zurich (LK).  

Technical support from Philips Healthcare, Best, The Netherlands, is gratefully acknowledged (LK, FP).

FK acknowledges grant support from the NIH under awards R01EB024532, P41EB017183 and R21EB027241.

OM acknowledges grant support from the Austrian Academy of Sciences under award DOC-Fellowship 24966.

\bibliography{main}

\clearpage
\listoffigures

\clearpage

\begin{table}
\centering
\resizebox{\textwidth}{!}{
\rowcolors{2}{gray!25}{gray!5}
\begin{tabular}{llllll}
\rowcolor{gray!55}
Submission &  Main Language & Licence & NUFFT/Gridding Type and Kernel interpolation& Sensitivities & CPU/GPU \\
University of California, Berkeley &  Python & MIT &SigPy's NUFFT / min-max~\citep{fessler2003} & SoS (Brain) / ESPIRiT~\citep{uecker2014} (Cardiac)  & both\\%
Berlin Ultrahigh Field Facility &  Matlab & MIT &BART  / Kaiser-Bessel \& Toeplitz emb.~\citep{jackson1991, wajer2001, uecker2010, ong2015}& SoS & both\\%
Eindhoven University of Technology&  Python & MIT &PyNUFFT / min-max~\citep{fessler2003} & None & both\\%
Swiss Federal Institute of Technology Zurich &  Matlab & TBD &In-house / Kaiser-Bessel - nearest~\citep{beatty2005} & SVD-Walsh~\citep{walsh2000} & CPU\\%
Karolinska Institutet &  Matlab & MIT & MIRT / min-max~\citep{fessler2003} & SoS & CPU\\%
Massachusetts General Hospital &  Matlab & n/a &BART  / Kaiser-Bessel \& Toeplitz emb.~\citep{jackson1991, wajer2001, uecker2010, ong2015} & ESPIRiT~\citep{uecker2014} & both\\%
New York University  &  Matlab & MIT & gpuNUFFT / Kaiser-Bessel - nearest~\citep{knoll2014} & Walsh~\citep{walsh2000} & both\\%
University of Southern California &  Matlab & MIT &In-house / Sinc & SoS & CPU\\%
Stanford University&  Python & MIT &In-house / Kaiser-Bessel - linear interp.~\citep{beatty2005} & SoS & CPU\\%
Graz, University of Technology, Institute of Computer Graphics and Vision &  Python & LGPL v3.0&gpuNUFFT / Kaiser-Bessel - nearest~\citep{knoll2014} & ESPIRiT~\citep{uecker2014} & GPU\\%
Graz, University of Technology, Institute of Medical Engineering &  Python & Apache-2 & In-house / Kaiser-Bessel - linear interp.~\citep{beatty2005} & ESPIRiT~\citep{uecker2014} & GPU\\%
Utah Center for Advanced Imaging Research, University of Utah &  Matlab & MIT &In-house / min-max~\citep{fessler2003} & Walsh~\citep{walsh2000} & both\\%
\end{tabular}}
\caption{Overview of the submissions and the main features used to create the results.}
\label{tab:algo}
\end{table}

\end{document}